\documentclass[11pt]{iopart}

\usepackage{graphicx}
\usepackage{subfigure}
\usepackage{float}
\usepackage{hyperref}
\usepackage[all]{hypcap}

\usepackage{placeins}

\begin{document}
	\title[Simulation of hydrogen pellets and SPI into RE beam in ITER]{Lagrangian particle simulation of hydrogen pellets and SPI into runaway electron beam in ITER}
	\author{ Shaohua Yuan$^1$, Nizar Naitlho$^1$, Roman Samulyak$^{1,2}$, Bernard P\'egouri\'e$^3$, Eric Nardon$^{3}$, Eric Hollmann$^4$, Paul Parks$^5$, Michael Lehnen$^6$
		\address{$^1$ Department of Applied Mathematics and Statistics, Stony Brook University, Stony Brook, NY, USA}
		\address{$^2$ Computational Science Initiative, Brookhaven National Laboratory, Uptown, NY, USA}
		\address{$^3$ CEA, IRFM, F-13108 Saint-Paul-lez-Durance, France}
		\address{$^4$ University of California - San Diego, La Jolla, CA, USA}
		\address{$^5$ General Atomics, San Diego, CA, USA}
		\address{$^6$ ITER Organization, Route de Vinon sur Verdon, 13115 St Paul Lez Durance, France}
		\ead{roman.samulyak@stonybrook.edu}
	}

\begin{abstract}
Numerical studies of the ablation of pellets and shattered pellet injection (SPI) fragments into a runaway electron beam in ITER have been performed using a time-dependent pellet ablation code [R. Samulyak at el., Nucl Fusion, 61 (4), 046007 (2021)]. The code resolves detailed ablation physics near pellet fragments and large-scale expansion of ablated clouds. The study of a single fragment ablation quantifies the influence of various factors, in particular the impact ionization by runaway electrons and cross-field transport models, on the dynamics of ablated plasma and its penetration into the runaway beam. Simulations of SPI performed using different numbers of pellet fragments study the formation and evolution of ablation clouds and their large-scale dynamics in ITER.  The penetration depth of ablation clouds is found to be of the order of 50 cm.
\end{abstract}

\maketitle

\section{Introduction}

Recent experimental results from DIII-D and JET \cite{Paz-Soldan2021,ReuxPaz-Soldan2021} suggest that pure D2 (or H2) shattered pellet injection (SPI) into a Runaway Electron (RE) beam may be a promising RE mitigation strategy for ITER. In DIII-D and JET, sufficiently large D2 injection is observed to cause rapid ($\sim$20 ms or less) thermal background plasma volume recombination and a simultaneous radial transport of background high-Z (argon) impurities out of the RE beam. Upon subsequent compression against the plasma wall, these “purged” RE beams are then observed to be lost to the wall in a single large MHD instability with large RE wetted area, low local heat loads, and nearly full conversion of RE current into thermal plasma current \cite{Paz-Soldan2021,ReuxPaz-Soldan2021}. Presently, the physics of this unusually large, single RE loss event is not understood. Experimentally, however, the large MHD event appears to depend on achieving background plasma volume recombination. In present devices, background plasma volume recombination is understood as resulting from rapid neutral diffusion of D and D2 into the vessel volume and subsequent molecular recombination of plasma ions \cite{HollmannBykov2020}. The initial D2 deposition profile from the pellet ablation does not appear to be important in present devices \cite{Shiraki2018}. In ITER, however, due to the larger plasma minor radius, it is expected that RE plateau diffusive timescales will be longer than in present devices \cite{HollmannBortolon2022}, so the initial pellet deposition profile may become important for determining whether or not plasma volume recombination can be achieved on timescales less than the RE plateau vertical drift time (~100 ms). Preliminary work based on a simple 1D model suggests that gas cloud expansion in the vicinity of SPI shards ablated by REs may improve injected D penetration into the RE beam \cite{Kiramov2020}. The main objective of this paper is to improve on this work and provide a numerical estimate of penetration of ablated mass into a RE beam resulting from the expanding SPI D2 shard ablation plumes.

Simulations of pellets and SPI using tokamak MHD codes are restricted to relatively coarse meshes that cannot resolve SPI fragments and fine physics details of the ablation process. In this work, we use a highly adaptive 3D Lagrangian particle pellet code \cite{SamulyakYuan2021} based on the Lagrangian particle method \cite{SamWangChen2018} for hydrodynamic equations. The code is capable of resolving a large spectrum of spatial scales ranging from sub-millimeter scales associated with phase transition processes on pellet surfaces and steep density gradients in cold ablation clouds to tokamak-scale expansion of ablation clouds.
By resolving relevant physics processes near pellet / fragment surfaces and in the far field, the LP code self-consistently computes pellet ablation rates and shows their dependence on 3D effects. The code was validated using experimental data on the injection of hydrogen fueling pellets and DIII-D experiment with small neon pellets  \cite{HollmannNaitlho2022}.

The paper is organized as follows. In Section \ref{sec:setup}, we describe the problem setup and main physics parameters. Main models, governing equations and their numerical implementation are presented in Section \ref{sec:models}. Results of numerical simulations of a single fragment injection and shattered pellet injection (SPI) into a runaway electron beam in ITER are presented in Section \ref{sec:simulation}. We conclude the paper with a summary of our results and plans for the future work.

\section{Problem setup}
\label{sec:setup}

Consider a large, cylindrical, pure hydrogen pellet with 28.5 mm diameter  and  57 mm length that is presently the foreseen size of large pellets in the ITER disruption mitigation system.
The pellet is injected with the velocity of $V=500$ m/s into a runaway electron beam in ITER. Before entering the plasma, the pellet is shattered into a large number of fragments creating a cloud with a 20 degree half angle and the injection time of about 2.2 ms. 

\begin{figure}[h!]	
	\centering
	\subfigure[Schematic of SPI into RE beam in ITER \label{schematic_a}]{\includegraphics[width=.8\textwidth]{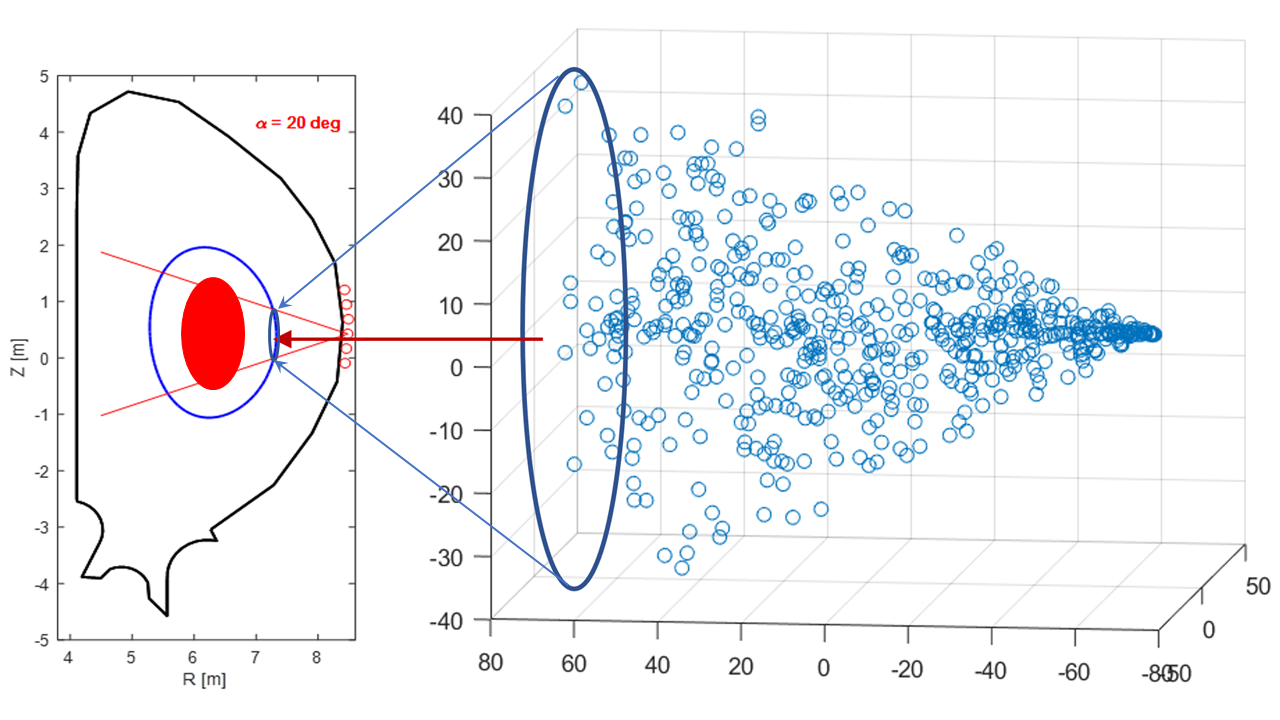}}
	\subfigure[Profile of relative RE density \label{schematic_b}]{\includegraphics[width=.7\textwidth]{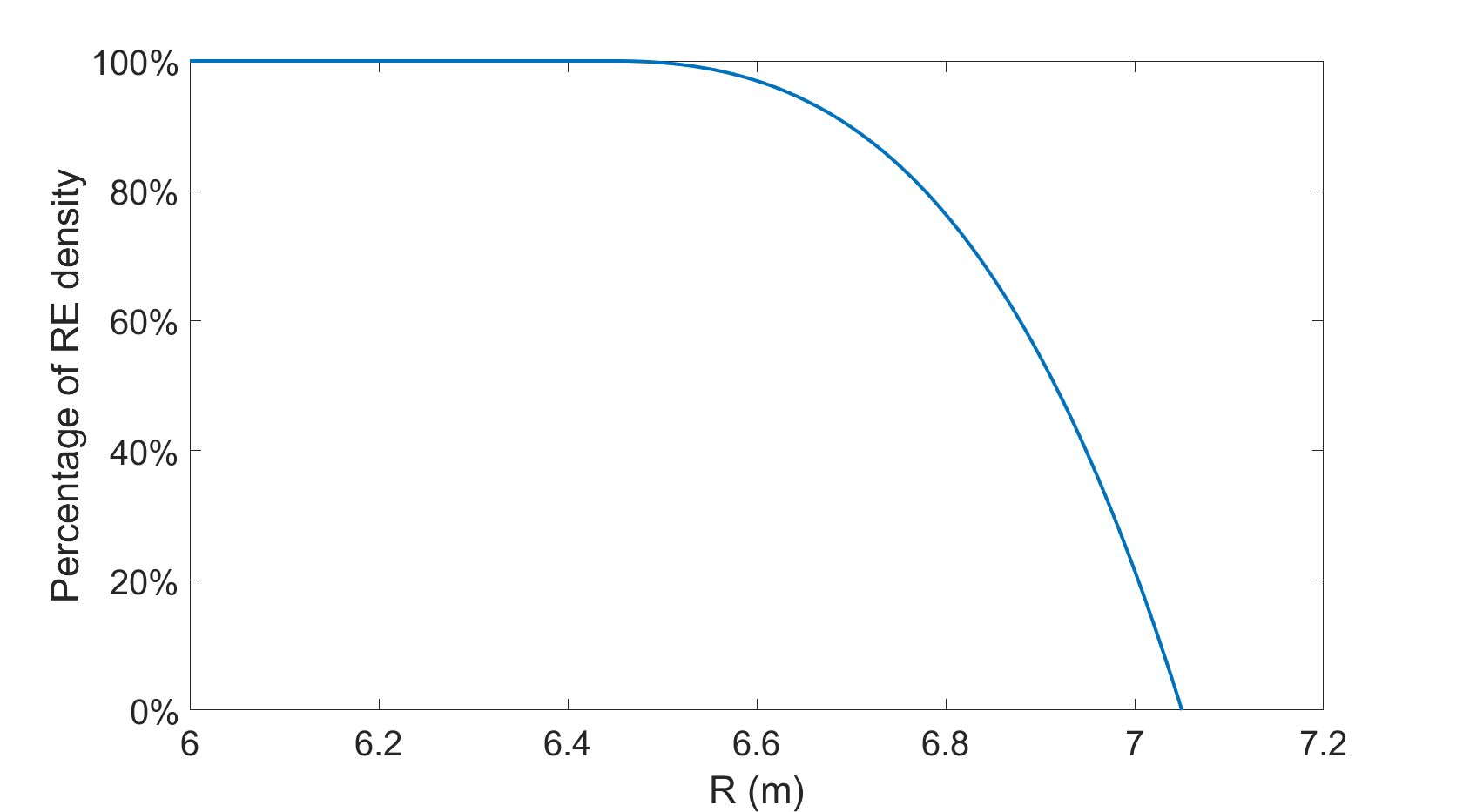}}
	\caption{(a) Schematic of SPI into RE beam in ITER. The left image shows ITER cross-section with a runaway electron beam region. The right image schematically shows enlarged distribution of SPI fragments.
	The injection direction is shown by the red arrow. (b) Profile of the relative RE density in the ramp between the RE core and the edge.}	
\label{SPI_schematic}
\end{figure}

Schematic of the simulation setup is shown in Figure \ref{SPI_schematic}. The left image in Figure \ref{schematic_a} shows an ITER cross-section containing a runaway electron beam region outlined by the blue line. The red ellipse denotes the core RE beam with constant RE density and the layer between the red and blue ellipses is an RE density ramp with 60 cm thickness. The right image in Figure \ref{schematic_a} shows an enlarged distribution of SPI fragments.
The real-size fragment distribution is injected into the RE region in the direction of the red arrow. Figure \ref{schematic_b} depicts the profile of relative (with respect to the core) RE density in the ramp between the RE core and the edge. The x-coordinate in this plot is consistent with the x-coordinate in the ITER cross-section plot.

The RE beam and its companion plasma parameters are approximated as follows: the RE beam is monoenergetic with $E_{RE}=20 MeV$ and the core of the RE beam has uniform density of $n_{RE}=5\times 10^{16}m^{-3}$ which corresponds to $10$ MA of runaway current. 
The companion plasma consists of a mixture of deuterium and neon with densities $n_{D}=1\times 10^{20}m^{-3}$ and $n_{Ne}=5\times 10^{19}m^{-3}$, where the neon density corresponds to pure Ne injection resulting in $50$ ms current quench time. Electron densities are based on coronal equilibrium ionization stages exclusive of ionization through the RE beam. The background plasma temperature is $T=20 eV$ and the electron density is $n_{e}=3\times10^{20}m^{-3}$. 

The goal of  this simulation study is to characterize the ablation of fragments, heating of the ablated material by the RE beam, and evolution of the ablated cloud as it ionizes, channels along magnetic fields lines and experiences cross-field transport by various drift forces. We are interested in computing the penetration depth of the ablated cloud into the RE beam and clarifying the dependence of ablation cloud evolution on the number and size of pellet fragments.   	
	
\section{Main models, governing equations and their numerical implementation}
\label{sec:models}

\subsection{MHD in low Magnetic Reynolds number approximation}

As a cryogenic pellet or an SPI fragment is injected into a cold tokamak plasma in the presence of a runaway beam, the pellet experiences an intense volumetric heating by REs and a rapid ablation (sublimation). A cold, neutral, high-pressure ablation cloud is formed around the pellet that initially expands spherically by pressure gradients. The expanding ablation material is heated up by REs, gradually ionizes via the thermal Saha mechanism as well as by direct impact of fast runaway electrons, and starts flowing predominantly along magnetic field lines by the Lorentz force action. To describe the evolution of such a cold, partially ionized ablation cloud, we use the system of MHD equations in the low Magnetic Reynolds number approximation, ${\delta B} / B \sim R_m << 1$, where $\delta B$ is the eddy current induced magnetic field. In Lagrangian coordinates, the equations are
\begin{eqnarray}\label{euler_eq}
	&&\frac{d \rho}{d t}  = - \rho \nabla \mathbf{u}, \\
	&&\rho \frac{d \mathbf{u}}{dt}= -\nabla P + \mathbf{J \times B} + \mathbf{f}_D, \label{eq:momentum}\\
	&&\rho \frac{de}{dt} = -P\nabla \cdot \mathbf{u} + \frac{1}{\sigma}\mathbf{J}^2 + Q_{RE},\label{eq:energy}\\
	&& P = P(\rho,e), \label{eq:eos}
\end{eqnarray}
where $d/dt$ is the Lagrangian time derivative, $\mathbf{u}$, $\rho$ and $e$ are the velocity, density and specific internal energy, respectively, $P$ is the pressure, $\mathbf{B}$ is the magnetic field induction, $\mathbf{J}$ is the current density, and $\sigma$ is the fluid conductivity. The runaway electron heat flux is represented by an external heat source $Q_{RE}$. Viscous forces in the ablation cloud are  neglected. As we show in Section \ref{sec:simulation}, the ablated material heating by cold background plasma via heat conduction is negligibly small in the energy balance in the ablation cloud, hence excluded in the equations. Finally, the ablation cloud experiences various drift forces in the transverse direction to  the magnetic field represented by the term $\mathbf{f}_D$. The corresponding forces are explained in detail in Section \ref{gradB}, dedicated to the cross-field transport.

The current density for (\ref{eq:momentum} - \ref{eq:energy}) is obtained from Ohm's law
\begin{equation}
	\mathbf{J}  = \sigma(-\nabla \phi + \mathbf{u \times B}),
\end{equation}
where $\phi$ is the electric potential in the cloud.  For a general 3D problem, the electric potential must be found from the following Poisson equation 
\[
\nabla \sigma\nabla \phi = \nabla \cdot (\mathbf{u \times B})
\]
that follows from the charge conservation equation $\nabla\cdot  \mathbf{J}  = 0$, subject to an appropriate boundary condition.  In this work, we assume that the ablation cloud is uniformly charged resulting in a constant value of $\phi$. The current density becomes
\[
\mathbf{J}  = \sigma_{\perp}\mathbf{u \times B}, 
\]
where ${\mathbf B} = (0,0,B_{\parallel})$ and the transverse conductivity $\sigma_{\perp}$ is approximated by a model derived for hydrogenic species in \cite{SamulyakLu2007}.
Similar approximations for ablation cloud were used in simulations of pellet ablation by hot thermal plasma electrons  \cite{SamulyakYuan2021,BosvielParks2021}

The EOS model used in this study implements two sources of ionization. First, it employs the local thermodynamic equilibrium (LTE) Saha model \cite{Zeldovich}. The second component of the model is based on the direct impact ionization of the ablated hydrogen by energetic REs. The electron impact ionization rates are computed as 
\begin{equation}
	R = N_e(r) \int v\sigma(E)f_e(E)dE ,
\end{equation}
where $v$ is the electron impact velocity, $\sigma(E)$ is the collision cross-section, $f_e$ is the electron energy distribution function, and $N_e(r)$ is the RE density that depends on the radial location in the tokamak as shown in Figure \ref{schematic_b}. Electron impact ionization rates for the parameter space of our problems of interest have been obtained courtesy of N. Garland \cite{NGarland}. 

We would like to comment on omitting hydrogen radiation losses in equation (\ref{eq:energy}). Using simulation data presented in later sections, we extracted time-dependence of hydrodynamic states on Lagrangian particles at various phases of the ablation cloud evolution and used the collisional-radiative code PrismSPECT \cite{BaileyRochau2009} to compute hydrogen radiation power density. The results show that hydrogen radiation power density is always much smaller compared to the RE heating power density. Only at the end of the process when the cloud density is very low, hydrogen radiation becomes of the same order but still smaller compared to the RE heating. The radiation of REs is always much higher compared to the radiation of hydrogen. Since a small fraction of radiation is absorbed by the hydrogen cloud, the RE radiation energy source partially compensates the hydrogen radiation energy sink. Since adding using full radiation hydrodynamics algorithms is not feasible for the present problems, neglecting radiation seems to be a reasonable 1st approximation. 

\subsection{Runaway electron heat flux}\label{RE_heat}

The energy of RE in our model is assumed to be in the range of 20 MeV. Consider an incident RE  beam passing through a pellet or ablated material with atomic density $n_{0}$ until it is stopped. The following expression \cite{breizman2019physics} gives the stopping power of the ultrarelativistic electrons:

\begin{equation}\label{eq:stoppower}
	\frac{d}{dt}\frac{\epsilon_{RE}}{m_{e}c^{2}}=-4\pi r^{2}_{e}c \ln \Lambda_{free}(n_{free}+\frac{1}{2}n_{bound}),
\end{equation}
where $\epsilon_{RE}$ is the RE energy, $r_{e}\equiv e^{2}/mc^{2}$ is the classic electron radius, $\ln \Lambda_{free} \approx 21$ is the Coulomb logarithm, and $n_{free}$ and $n_{bound}$ are the free and the bound electron number densities, respectively. In the solid state, nearly all the electrons are bound so we have $n_{free}=0$ and $n_{bound}=Zn_{0}$, where $Z$ is the atomic number of the pellet material. Equation (\ref{eq:stoppower}) gives the stopping length of RE \cite{Kiramov2020} as
\begin{equation}
	L \approx \frac{\epsilon_{RE}}{m_{e}c^{2}}\frac{1}{2\pi r_{e}^{2}\ln \Lambda Zn_{0}}.
\end{equation}
For the solid hydrogen density $n_{0}=4.5\times10^{28}\, 1/m^{-3}$, the stopping length of RE is $L_{H}\approx78$ cm. Therefore, we conclude that REs pass through the pellet / ablated clouds multiple times before being stopped and provide volumetric heating.  The heat source $Q_{RE}$ from REs streaming into the pellet and the ablation cloud can be computed from (\ref{eq:stoppower}) as
\begin{equation}\label{RE_eq}
	Q_{RE} = 4\pi m_{e}r^{2}_{e}c^{3} n_{RE}\ln \Lambda_{free}(n_{free}+\frac{1}{2}n_{bound})
\end{equation}

\subsection{Transverse flow dynamics}\label{gradB}
\label{sec:transverse_dynamics}
	
A cold and weakly ionized ablation cloud polarizes as it travels across magnetic field lines and the resulting polarization $\mathbf{E \times B}$ drift attempts to maintain the cloud motion with the initial injection velocity. In addition, the cloud polarization evolves, due to (a) the cloud-plasma pressure gradient which tends to maintain it and (b) the Alfvén wave emission and internal and external cloud self-connections along field lines which tend to reduce it \cite{Koechl2012}. The resulting equation governing the $\mathbf{E \times B}$ drift velocity is:

\begin{eqnarray}\label{eq:drift}
&&	\frac{dv_D}{dt} = \frac1{1+(1-P_{Alf} - P_{con}) (L'_{con}\rho_{\infty} / Z_0 \rho )}\times \\
&&	\left[ 	\frac{qR}{Z_0}\sin\left(\frac{Z_0}{qR}\right)\frac{2 <P(1+\frac{M_\parallel^2}{2}) - P_\infty>}{R<\rho>}  
- v_D\frac{B_{||}^2}{<\rho>} H(\tau_{Bdiff} - \tau_{adv})\times  \right. \nonumber \\
&&\left. \left\{P_{Alf} \frac2{\mu_0 v_A} + 
	P_{con}\frac{\left[ 1-\exp(-t/(\tau_{coll}^e + \tau_L)) \right] \pi r_0^2\-\sigma_{\infty}}{L_{con}} \right\}	\right]. \nonumber
\end{eqnarray}
In this expression,

$< A > \equiv \int_{-Z_0}^{Z_0} Adz$, the integral is along a magnetic field line, $A$ is any quantity, 

$Z_0$: half-cloud length, 

$v_D$: drift velocity,

$P_{con}$: proportion of cloud section self-connected (+ vs. -) at time $t$,

$P_{Alf}$: proportion of cloud section not self-connected at time $t$,

$L'_{con}$: arithmetic average of the length of the field lines self-connecting the cloud,

$L_{con}$: harmonic average of the length of the field lines self-connecting the cloud,

$B_{||}$: background magnetic field,

$\rho$: cloud density,

$P$: cloud pressure,

$\rho_{\infty}$: background plasma density,

$P_{\infty}$: background plasma pressure,

$r_0$: cloud radius, 

$q = 2$: tokamak safety factor, 

$R$: tokamak major radius,

$M_\parallel$: Mach number of the plasmoid parallel expansion,

$\sigma_{\infty}$: plasma parallel conductivity,

$\tau_{coll}^e$: plasma electron collision time,

$\tau_L = \sigma_{\infty} \mu_0 P_{con}r_0^2$: damping current delay time,

$H$: the Heaviside function that depends on

$\tau_{Bdiff} = \mu_0 \sigma_{\perp} L_{char}^2$: characteristic time for magnetic field diffusion over characteristic length $L_{char}=5$ cm,

$\tau_{adv} = L_{char}/v_D$: characteristic time for cloud motion across magnetic field lines.

After the ablation cloud ionizes and expands along magnetic field lines, it starts drifting across the magnetic field in the direction of the outward major radius $R$. This motion, described by the left term in the second row of equation (\ref{eq:drift}), has been attributed to $\mathbf{E \times B}$ drift resulting from a vertical curvature and grad-B drift polarization induced inside the ionized ablated material by the $1/R$ toroidal magnetic field variation \cite{Baylor_2000, Parks_2005,Matsuyama2012}. As the cloud continues to heat-up and ionize, the characteristic magnetic field diffusion time $\tau_{Bdiff}$ becomes longer compared to the 
characteristic time for cloud motion across magnetic field lines, $\tau_{adv}$, i.e. magnetic flux becomes frozen into the cloud. At this time, the Alfven wave drag and the external connection (P\'egouri\'e) current - induced drag, described by the two terms in the third line of equation (\ref{eq:drift}), become essential \cite{Koechl2012,Commaux2010}.  Formally, these terms are added as the Heaviside function in the second line of equation (\ref{eq:drift}) changes from zero to one. As we mentioned above, the plasmoid polarizes in the transverse direction as it moves across the magnetic field lines. This creates a rapidly expanding Alfven-wave induced polarization in the ambient plasma. After time $\tau_{con} = L'_{con}/v_A$, the plasmoid becomes self-connected by this polarization perturbation.   Proportion of the cloud section not self-connected at this time, $P_{Alf}$, contributes to the Alfven wave drag force. Proportion of  the cloud section connected to the region of opposite polarity, $P_{con}$, contributes to the external connection current drag term. Proportion of the cloud  section connected to the region of the same  polarity, $P_{same} = 1- P_{con} - P_{Alf}$, increases the effective moving mass as indicated in the first term on the right-hand-side in equation (\ref{eq:drift}). Further details, including the topology of self-connection paths in a tokamak, can be found in \cite{Koechl2012,Commaux2010}.

\subsection{Numerical implementation}

The system of equations (\ref{euler_eq}) - (\ref{eq:eos}) is solved by the Lagrangian Particle (LP) pellet ablation code \cite{SamulyakYuan2021} based on the Lagrangian particle method \cite{SamWangChen2018} for hydrodynamic equations. 
The code is applicable to pellets and SPI fragments ablated by hot plasma and REs.  
The LP method represents fluid cells with Lagrangian particles and is suitable for the simulation of complex free surface or multiphase flows. It is highly adaptive to density changes, a critically important property for 3D simulations of the ablation of pellets and, especially, SPI fragments.
The LP-based pellet ablation code implements MHD equations in the low magnetic Reynolds number approximation and the physics models described above. The LP code is optimized for massively parallel supercomputers by using p4est ("parallel forest of K-trees" ) \cite{BursteddeWilcoxGhattas11}, a parallel library that implements a dynamic management of a collection of adaptive K-trees on distributed memory supercomputers. 

Lagranigan particle algorithms for pellet / SPI fragment ablation by hot plasma electrons were developed in  \cite{SamulyakYuan2021} and validated using experimental data on the injection of hydrogen fueling pellets and DIII-D experiment with small neon pellets \cite{HollmannNaitlho2022}. 
Physics models for pellet ablation clouds in REs beams were implemented in LP within the present work. In addition, the cross-field transport model for ablated clouds has been significantly improved. Since   \cite{SamulyakYuan2021} dealt with relatively small geometric domains around ablating pellets, sufficient for computing pellet ablation rates, it implemented simplified grad-B drift and Alfven wave drag models responsible for establishing pellet shielding lengths. The improved model described in Section \ref{sec:transverse_dynamics}
is necessary for large-scale simulation of ablated clouds reported in this work.

\section{Simulation results}\label{sec:simulation}

\subsection{The simulation of a single fragment into RE beam}

For better understanding of main processes in the ablation cloud, we simulate first a single  $5 mm$ diameter hydrogen pellet injected with initial velocity of $500m/s$ into the runaway beam from the low-field-side, as shown in the schematic in Section \ref{sec:setup}. At the start of the simulation, the pellet is located at the RE beam edge since the pellet ablation in low temperature plasma is negligibly small: the loss of pellet mass due to the ablation by thermal electrons was estimated to be 2 \%.

\begin{figure}[H]
\centering
\subfigure[Density (1/cc) at  $t=40\mu s$]{\includegraphics[width=.49\linewidth]{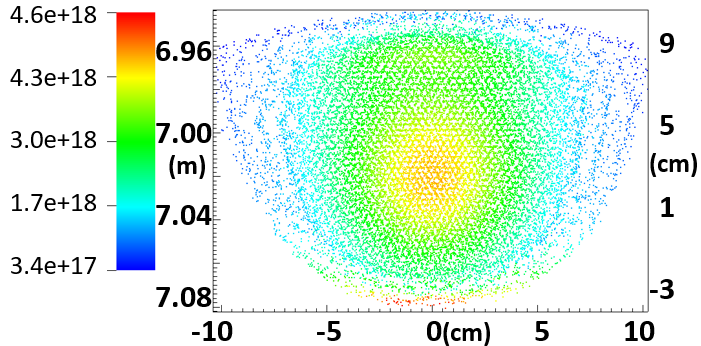}}
\subfigure[Density (1/cc) at  $t=80\mu s$]{\includegraphics[width=.49\linewidth]{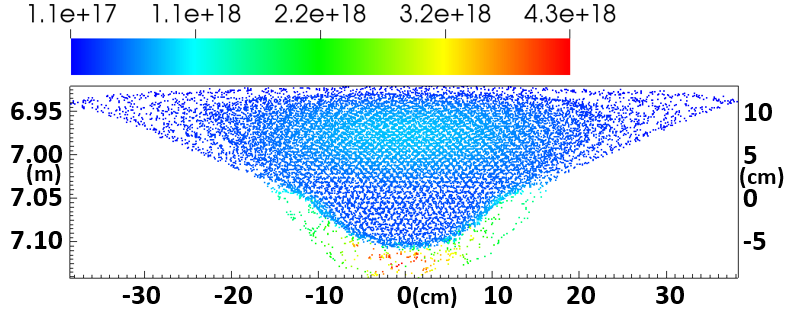}}
\subfigure[Pressure (bar) at  $t=40\mu s$]{\includegraphics[width=.49\linewidth]{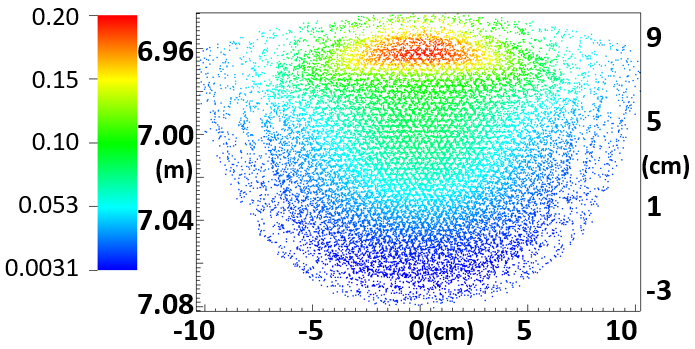}}
\subfigure[Pressure (bar) at  $t=80\mu s$]{\includegraphics[width=.49\linewidth]{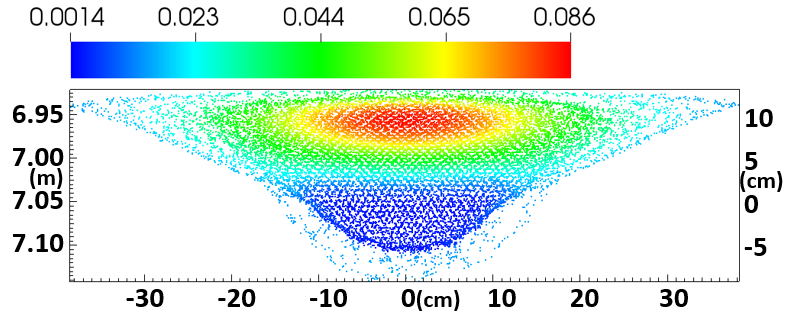}}
\subfigure[Temperature (eV) at  $t=40\mu s$]{\includegraphics[width=.49\linewidth]{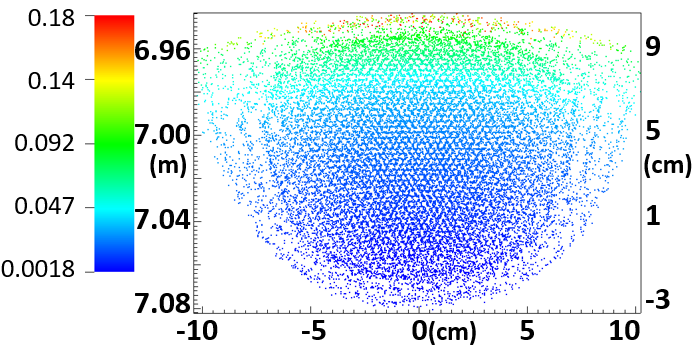}}
\subfigure[Temperature (eV) at  $t=80\mu s$]{\includegraphics[width=.49\linewidth]{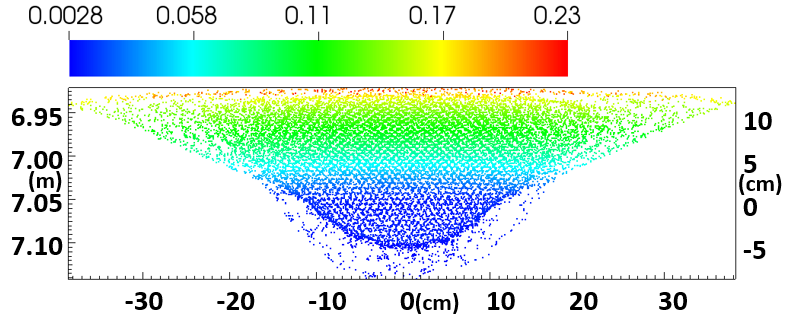}}
	\caption{Distribution of ablation cloud states on 2D slices at $t=40\mu s$ (left images) and $t=80\mu s$ (right images). Horizontal axis is along magnetic filed lines. There are two vertical scales on both sides of images: penetration depth (right axis), global coordinates (left axis). RE beam region is above 70 cm in vertical direction. The pellet is injected from the bottom side of the plot.}
	\label{singlepelletearlyprofile}
\end{figure}

As the pellet enters the RE beam ramp, it rapidly sublimates and forms a cold, dense ablation cloud. The short-time-scale sublimation process is not explicitly resolved - the pellet is converted instantaneously into gas state at the RE beam edge in our numerical model. 
Figures \ref{singlepelletearlyprofile}(a)-(f) show states of the ablation cloud in early stages of expansion, at 40$\mu s$ (left images)  and 80$\mu s$ (right images). The expansion at 40$\mu s$ remains almost spherical, with some deformation at the top due to extra heating in the non-uniform RE ramp. Ionization and MHD effects become visible at 80$\mu s$ as the top part of the cloud expands along magnetic field lines and experiences the grad-B drift.

\begin{figure}[H]
	\centering
	\subfigure[Connection coefficients]{\includegraphics[width=.49\linewidth]{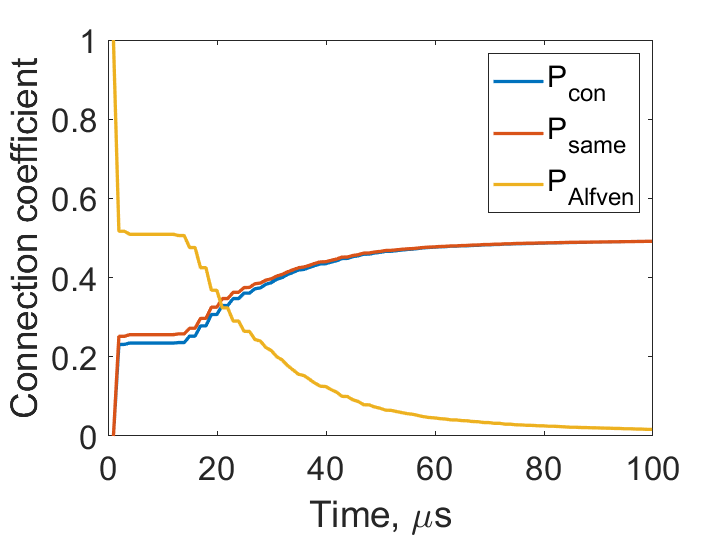}}
	\subfigure[Ratio of drift terms]{\includegraphics[width=0.5\linewidth]{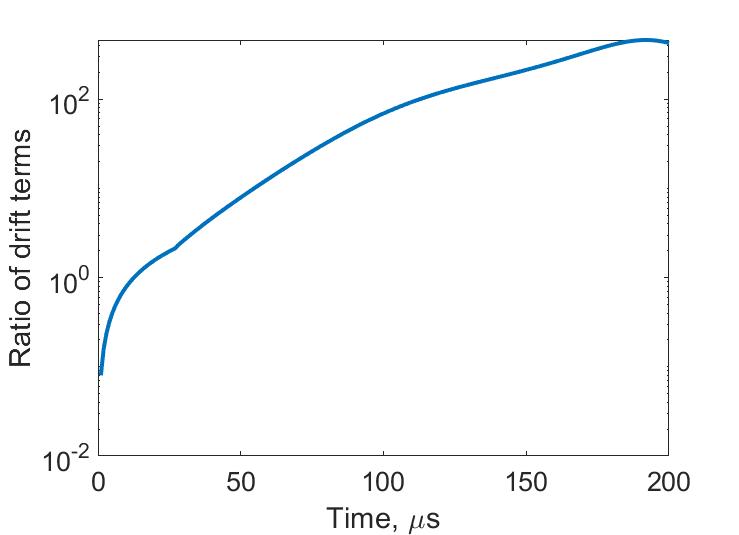}}
	\caption{(a) Evolution of connection coefficients $P_{con}$, $P_{Alf}$, and $P_{same}$. (b) Ratio of magnitudes of the external connection damping term and the Alfven wave drag term.}
	\label{drag_forces}
\end{figure}

\begin{figure}[H]
	\centering
	\subfigure[Drift velocity (m/s) at 0.75 ms]
	{\includegraphics[width=1.0\linewidth]{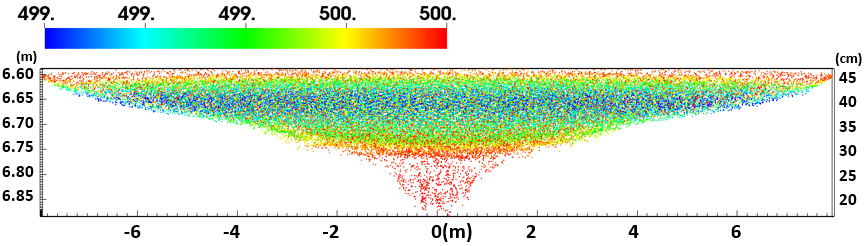}\label{driftv_0.75}}
	\subfigure[Drift velocity (m/s) at 1 ms]
	{\includegraphics[width=1.0\linewidth]{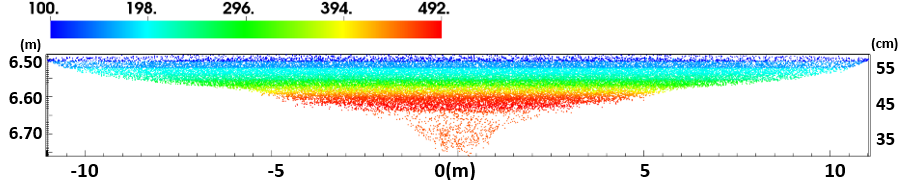}\label{driftv_1}}
	\subfigure[Drift velocity (m/s) at 1.05 ms]
	{\includegraphics[width=1.0\linewidth]{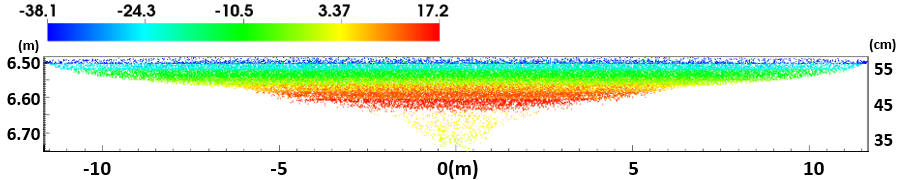}\label{driftv_105}}
	\caption{Distribution of the ablation cloud drift velocity at 0.75ms, 1ms and 1.05ms. The longitudinal axis (m) is along magnetic field lines. There are two vertical axes: the left axis (m) is the global coordinate and the right axis (cm) is the penetration depth.  The pellet is injected from the bottom side of the plot. The positive direction of drift velocity is the pellet injection direction.}
	\label{driftevolve}
\end{figure}

\begin{figure}[H]
	\centering
	\subfigure[Penetration depth]{\includegraphics[width=0.49\linewidth]{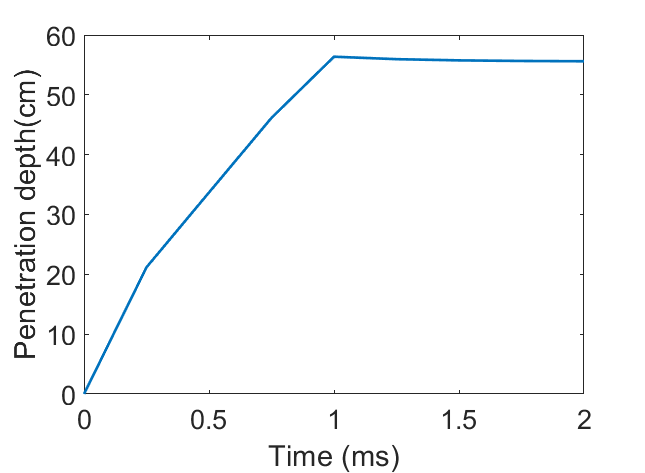}\label{penetration}}
	\subfigure[Cloud length]{\includegraphics[width=0.49\linewidth]{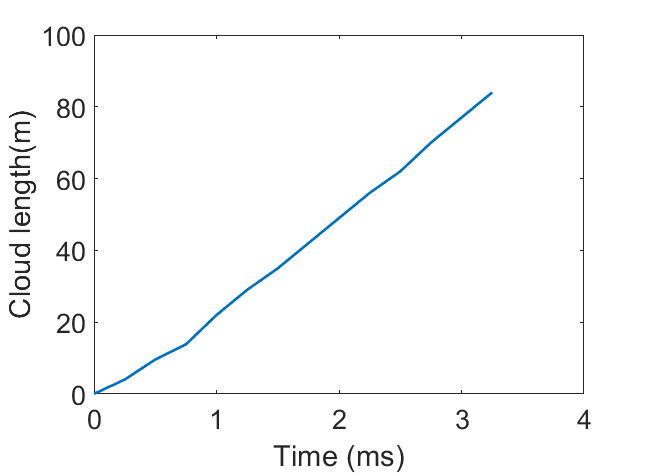}\label{cloud_length}}
	\caption{(a) Evolution of the penetration depth of the ablation cloud into RE beam. (b) Evolution of the ablation cloud length. }
	\label{cloudlength}
\end{figure}

As the cloud heats-up and ionizes, the magnetic field diffusion time exceeds the cloud advection time and the Alfven wave and external connection current drag terms turn on by the Heaviside function in equation (\ref{eq:drift}). Figure \ref{drag_forces} explains the dynamics and relative magnitude of these terms. During the first 15 microseconds of the action of drag terms, the magnitudes  of both connection coefficients  are close, $P_{con} \sim 0.3$ and $P_{Alf} \sim 0.37$, and the Alfven wave drag dominates. The Alfven wave drag connection coefficient reduces to small values within the next 100$\mu s$ and the external connection current drag becomes the dominant factor. These terms rapidly change the penetration dynamics of the ablation cloud as shown in Figure \ref{driftevolve}. At 0.75 ms after injection, the whole cloud still moves with the transverse velocity close to the injection velocity of 500 m/s (Figure \ref{driftv_0.75}). This is explained by a very low ionization level of the ablation cloud at early time resulting in a weak grad-B drift force. Ionization of front layers in the ablation cloud increases after 0.75 ms and the grad-B drift significantly slows them down. Figure \ref{driftv_1} shows  that top layers of the cloud slow down by 5 times while the bottom layers are still largely unaffected at 1 ms. At this time, the Heaviside function turns on and the cloud starts experiencing the Alfven wave and the external connection current drag forces. The penetration rapidly stops within the next 50 $\mu s$ as we see in  Figure \ref{driftv_105}: the top layers slowly drift backward by the grad-B drift while the bottom layers very slowly continue the forward motion. The penetration depth evolution is shown in Figure \ref{penetration}. We would like to note that we turned on the Heaviside function at exactly 1 ms for simplicity of analysis; this is within ~ 4\% from the exact timing when $\tau_{Bdiff} = \tau_{adv}$. Due to a small diameter of the ablation cloud, the Heaviside function turns on the drift forces uniformly across the cloud. We employ a more realistic approximation for SPI simulations in the next section.

\begin{figure}[H]
	\centering
	\includegraphics[width=.6\linewidth]{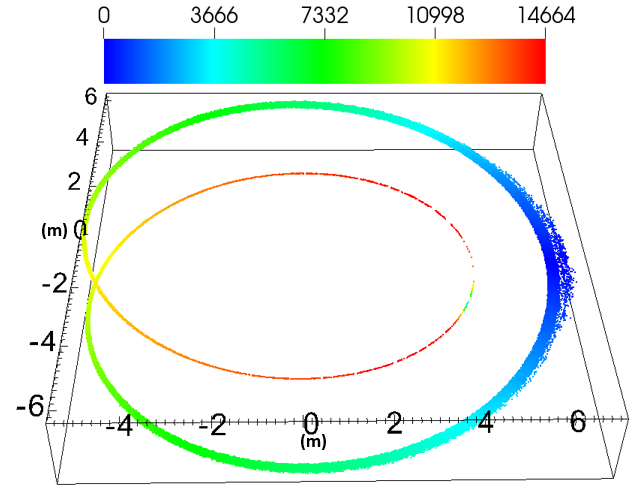}
	\caption{Long-term dynamics of the ablation cloud. Velocity distribution (m/s) is shown.}
	\label{cloud_q2}
\end{figure}

The longitudinal cloud expansion remains almost linear in time throughout simulation (Figure \ref{cloud_length}). Since the drag terms also prevent strong grad-B drift, the cloud remains at approximately the same depth in the RE beam. At about 3 ms, the cloud reconnects longitudinally at the safety factor $q=2$ (Figure \ref{cloud_q2}).

\begin{figure}[H]
\subfigure[Saha ionization]{\includegraphics[width=.5\linewidth]{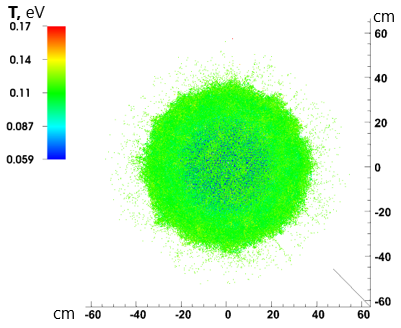}\label{thermalonly}}
\subfigure[Saha + impact ionization]{\includegraphics[width=.5\linewidth]{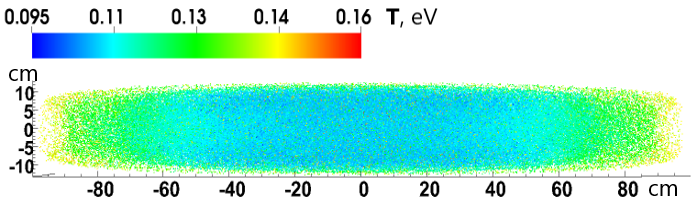}\label{thermalsaha}}
\caption{Temperature distribution (eV) of the ablation cloud at 0.2ms with (a) thermal Saha ionization and (b) both Saha ionization and ionization by impact.}
\label{ionizationbyimpact}
\end{figure}

Since our simulations used the direct electron impact ionization in addition to the thermal Saha ionization, it is important to clarify the role of the two ionization processes. We do this in simplified geometry and physics approximations to avoid the influence of other factors. In particular, we artificially turn off the cross-field terms (\ref{eq:drift}) and place the pellet in a uniform RE beam. In 1MA RE current, the cloud expands spherically and no MHD effects are observed if only thermal Saha  ionization is used. At $0.2ms$, the ionization level in the bulk of the cloud is around $10^{-7}$ (Figure \ref{thermalonly}). After adding the electron ionization by impact, we observed significant MHD effects in the ablation cloud and the channeling of the ablated material along magnetic field lines. The ionization level in the bulk of the cloud was around $0.004$ (Figure \ref{thermalsaha})  and the cloud length reached $190cm$ after $0.2ms$. If the RE current is increased to 10 MA, the Saha ionization alone is sufficient to cause the longitudinal motion along magnetic field lines, but the addition of impact ionization changes quantitative characteristics of the ablation cloud dynamics. At 200$\mu s$, the cloud was 3.2 m long and 40 cm in diameter 
with only Saha ionization and it expanded to 6.6 m length and reduced to 20 cm diameter when the impact ionization was added. Therefore, the electron impact ionization changes the cloud dynamics at all conditions, and it is especially important at low values of the RE density in the ramp.  

We would like to add a note on the heat diffusion in the ablation cloud.  Our simulations show that the longitudinal heat diffusion provides negligibly small factor in the energy balance. Even for the most favorable case for the longitudinal heat diffusion - 20 eV background plasma, only small regions at the front and back of the ablation cloud heat up to the background temperature at times relevant to our simulations; the bulk ablation plasma temperatures are mostly defined by the RE energy deposition and the expansion dynamics remain practically the same. Since the transverse heat conductivity is several orders of magnitude lower than the parallel conductivity, the transverse heat diffusion does not have any noticeable effect on the cloud dynamics despite the difference in scales: the cloud radius is much smaller than the cloud length. We verified our results by comparing 3D simulations to simplified 1D simulations. 
Considering the small effect of the thermal diffusion, 3D simulations reported in this work were performed with the heat diffusion turned off to reduce computing time. This explains why the temperature of the ablated plasma edge is not at the background plasma temperature.

\subsection{Simulation of SPI into RE beam}

In this section, we report numerical simulation results of SPI into the RE beam in ITER. We assume uniform size of pellet fragments and compare simulations using 555 fragments of 5 mm diameter and 5000 fragments of 2.4 mm diameter. Each cloud of SPI fragments forms a cone with a 20 degree semi-angle. The distribution of fragments is uniform with random perturbations. The injection time for clouds of both fragment sizes is 2.2 ms. 

\begin{figure}[H]
	\subfigure[555 fragments]{\includegraphics[width=.5\linewidth]{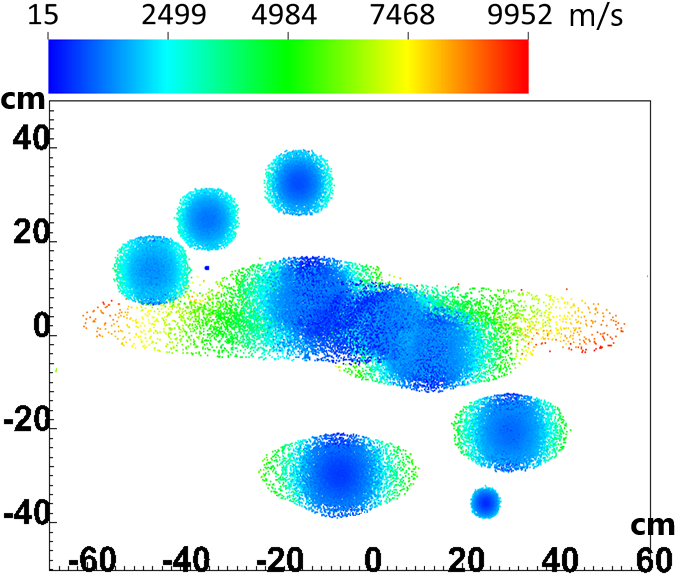}\label{vel_100mus_555}}
	\subfigure[5000 fragments]{\includegraphics[width=.5\linewidth]{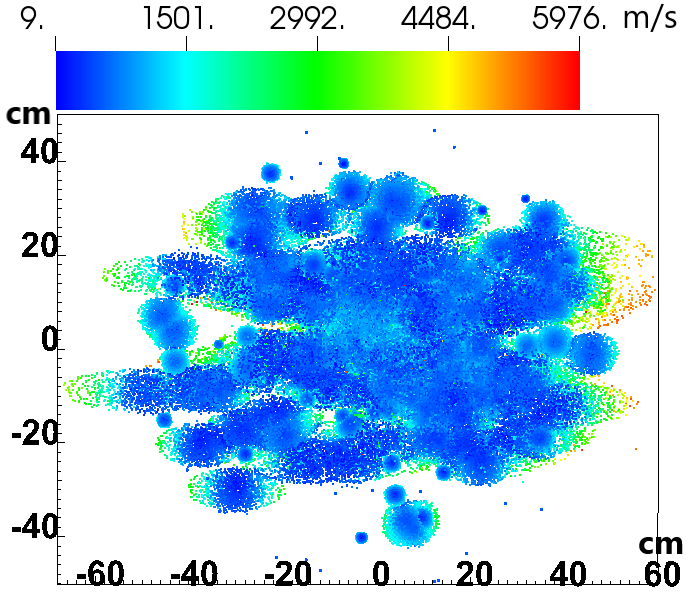}\label{vel_100mus_5000}}
	\caption{Initial phase of SPI: state of fragments after 100 $\mu s$. Velocity distribution (m/s) in ablated clouds is shown in the injection direction (view from the LFS) for (a) 555 fragments and (b) 5000 fragments.}
	\label{vel_100mus_spi}
\end{figure}

As first fragments enter the RE beam ramp, they rapidly ablate and form dense, high-pressure individual clouds (called here cloudlets). The initial stage of the injection and ablation process (at 100 $\mu s$) is illustrated in Figure \ref{vel_100mus_spi} which shows the ablated material velocity distribution in the direction of injection. At this time, we observe  small spherical cloudlets created from most recent fragments as well as larger cloudlets from earlier fragments that became sufficiently ionized and expanded along magnetic field lines by MHD forces. Cloudlets from larger size fragments (Figure \ref{vel_100mus_555}) evolve individually at this stage while cloudlets from smaller fragments (Figure \ref{vel_100mus_5000}) merge into a global cloud. In addition, higher initial pressure in cloudlets from 5 mm diameter fragments push some fraction of  cold and neutral ablated material away from the RE region. Similar process is also present for smaller SPI fragments but to a less extent, resulting in more material being deposited in the RE region at earlier stages of the ablation process. At later time, cloudlets created by large fragments also merge into a single albeit less uniform ablation cloud.

The ablation clouds continue to heat-up by absorbing the energy of REs, expand along magnetic fields lines, penetrate into the RE beam ramp by the $E\times B$ drift, and partially drift in the direction of large tokamak radius by the grad-B drift force. At some point in time, the front layers of the clouds in the injection direction become sufficiently ionized that the magnetic field diffusion time exceeds the advection time, causing the Alfven wave drag and external connection currents forces. Because of a large transverse size of the ablation clouds, we do not turn-on the drag forces uniformly in the entire cloud, as we did in simulations of a single fragment, but instead check  the Heaviside function argument in equation (\ref{eq:drift}) individually for small tubes of the ablated material along magnetic field lines and turn-on the drag forces accordingly. These forces decelerate the clouds and prevent them from penetrating deeply into the RE region.

\begin{figure}[H]
	\centering
	\subfigure[555 fragments]{\includegraphics[width=.80\linewidth]{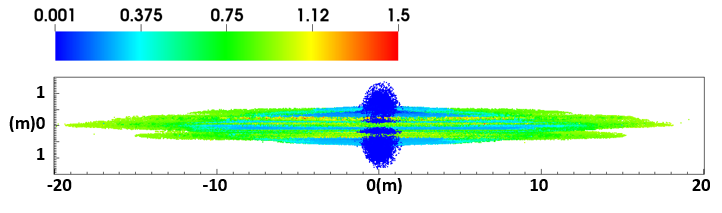}\label{T_1ms_555}}
	\subfigure[5000 fragments]{\includegraphics[width=.80\linewidth]{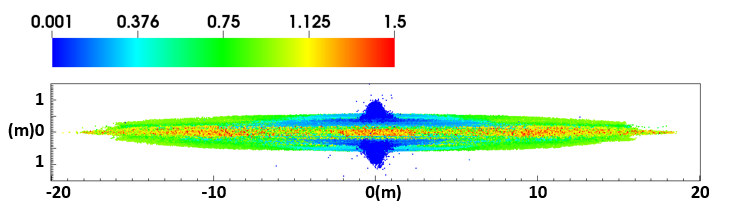}\label{T_1ms_5000}}
	\caption{State of SPI ablation cloud near maximum penetration depth, at time 1 ms. Temperature distribution (eV) in ablated clouds is shown in the injection direction  (view from the HFS) for (a) 555 fragments and (b) 5000 fragments.}
	\label{T_1ms_spi}
\end{figure}

\begin{figure}[H]
	\subfigure[555 fragments]{\includegraphics[width=.55\linewidth]{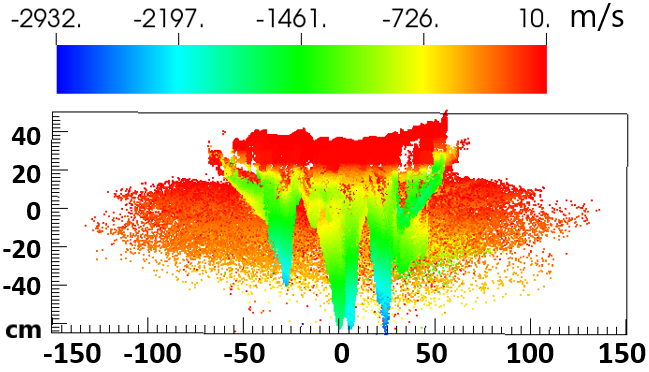}\label{transverse_vel_1ms_555}}
	\subfigure[5000 fragments]{\includegraphics[width=.45\linewidth]{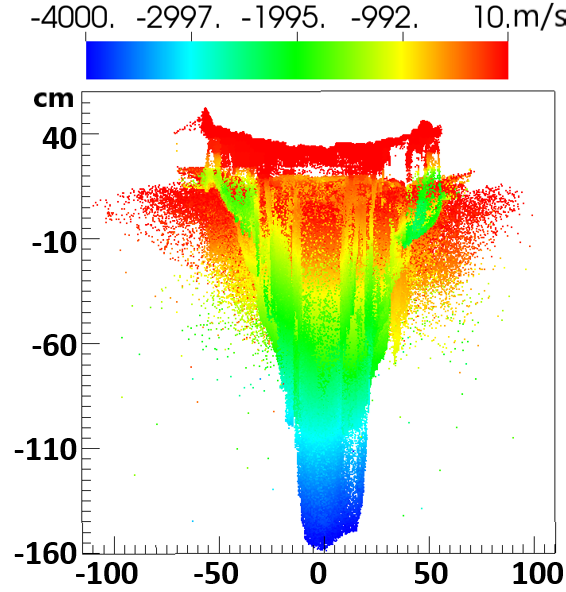}\label{transverse_vel_1ms_5000}}
	\caption{Transverse velocity distribution (m/s) in ablated clouds for (a) 555 fragments and (b) 5000 fragments near maximum penetration depth at time 1 ms. The line of view is along magnetic field lines and the vertical axis is along the injection direction, giving the penetration depth.}
	\label{transverse_vel_1ms_spi}
\end{figure}

Figures \ref{T_1ms_spi} shows temperature distributions in both clouds at the time of their maximum penetration into the RE beam. Plots show that the longitudinal expansion (approximately 36 m) and temperature values are very similar for both clouds obtained from large and small size fragments. Close to the pellet fragment injection region, we observe cold clouds of neutral material, approximately spherical in shape, expelled from the RE beam. The fraction of expelled cold particles is about 15\% and 10\% for the 555 and 5000 fragment SPI, respectively. At this time, a significant fraction of hot ablated material has also drifted outside of the RE region: the corresponding fractions for the 555 and 5000 fragment SPI are 50\% and 65\%. Depositing more ablated material into the RE beam at the initial stages of ablation in the 5000 fragment SPI leads to higher pressure values in the ablation cloud and stronger grad-B drift (Figure \ref{transverse_vel_1ms_5000}). As a result, less material remains in the RE beam over longer time.

\begin{figure}[H]
	\centering
	\subfigure[555 fragments]{\includegraphics[width=.6\linewidth]{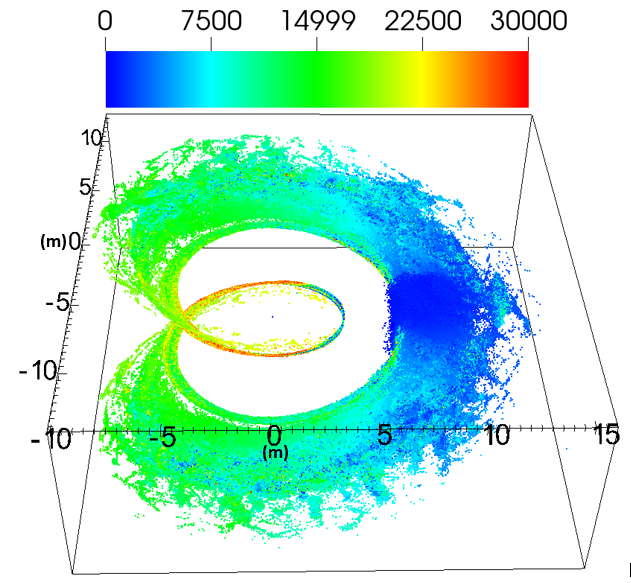}\label{vel_global_555}}
	\subfigure[5000 fragments]{\includegraphics[width=.6\linewidth]{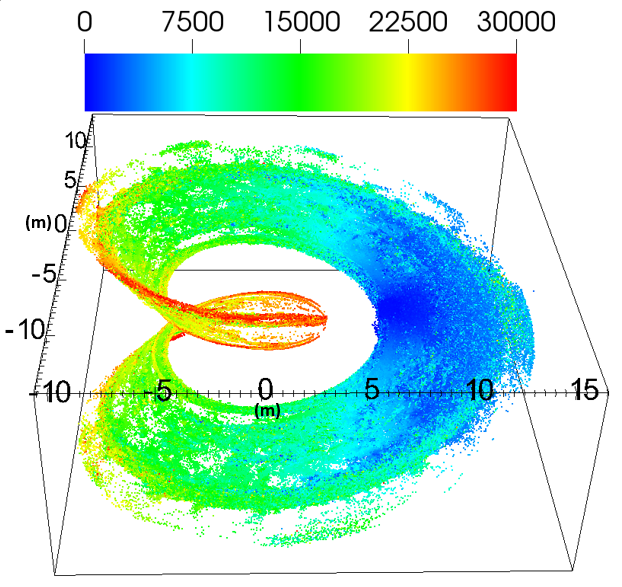}\label{vel_global_5000}}
	\caption{State of SPI ablation cloud after longitudinal reconnection at $q=2$ at time 1.7 ms. Velocity distribution (m/s) in ablated clouds is shown for (a) 555 fragments and (b) 5000 fragments.}
	\label{vel_global_spi}
\end{figure}

The global dynamics of ablation clouds is shown in Figure \ref{vel_global_spi}. At approximately 1.7 ms, the clouds re-connect in the longitudinal direction assuming the tokamak safety factor of 2. Both clouds contain wide layers of ablated hydrogen drifting beyond the RE region. As new fragments are still injected into the RE beam, the fraction of material drifted outside of the RE region slowly increases: it is a few percent higher at the re-connection time compared to the reported values at 1.1 ms.  

\begin{figure}[H]
	\centering
	\includegraphics[width=.7\linewidth]{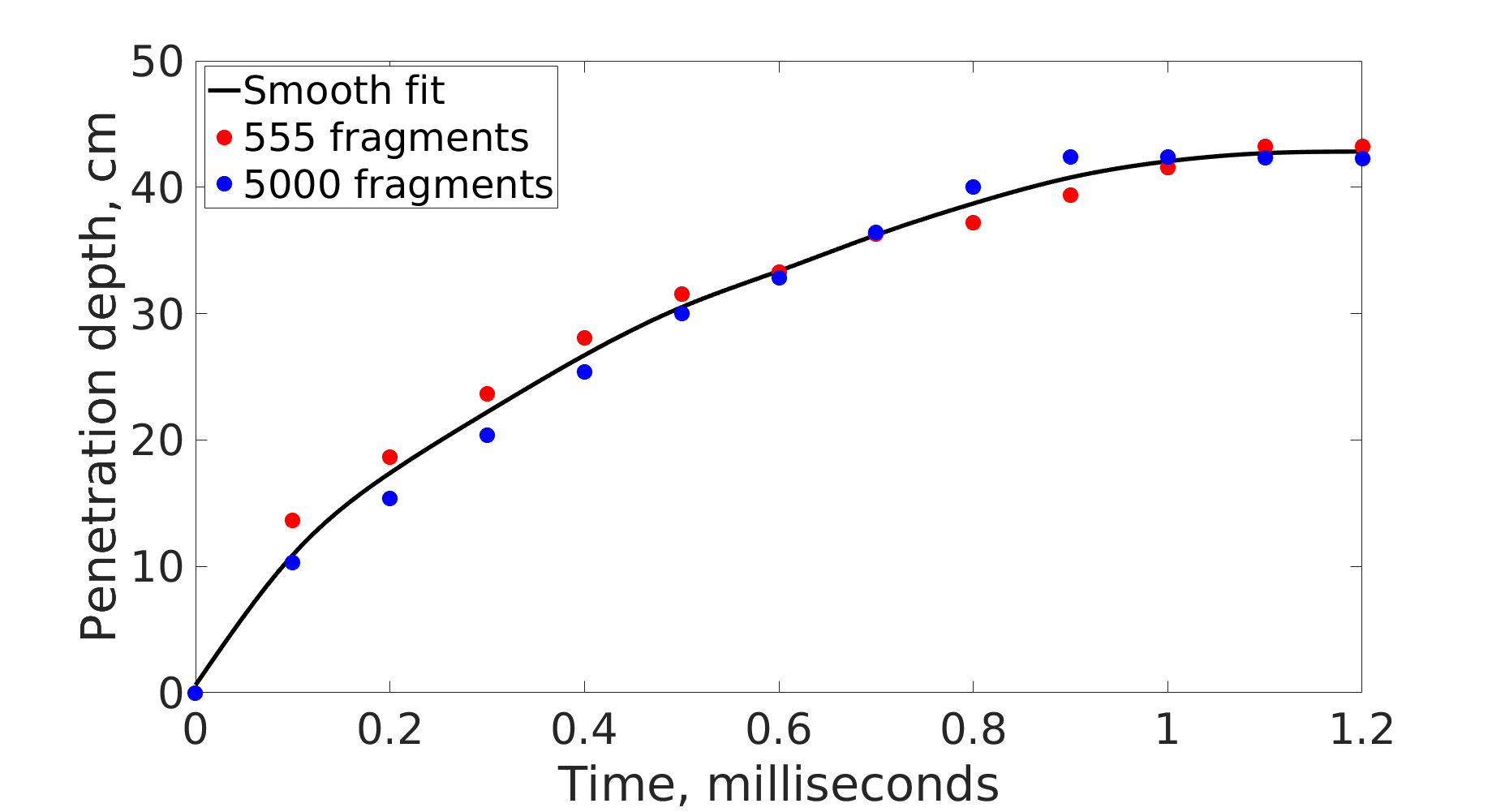}
	\caption{Evolution of penetration depth of SPI ablated cloud into the runaway beam.}
	\label{penetration_spi}	
\end{figure}

The penetration depth of the leading edge of both clouds in the RE beam is similar; each case is plotted with different color dots in Figure \ref{penetration_spi} and the line corresponds to a polynomial fit using both sets of data. 
The penetration depth is slightly smaller compared to that of the single cloud fragment reported in the previous section.

\section{Conclusions and future work}

Lagrangian particle simulation of the evolution of ablation clouds of hydrogen pellets and SPI fragments heated by a runaway electron beam in ITER have been performed. To accomplish this task, capabilities of the Lagrangian particle pellet / SPI code \cite{SamulyakYuan2021} have been extended by including physics models for the injection of hydrogen pellets and SPI into a runaway electron beam. In particular, ionization of ablated materials by direct electron impact and the volumetric heating by REs have been implemented. The transverse flow model has been significantly improved by including the internal (poloidal) and external P\'egouri\'e connection currents and the corresponding numerical algorithms have been developed and implemented. 

Simulations of the ablation of a single pellet fragment in the RE region in ITER demonstrated that the thermal Saha ionization alone is not sufficient for ionizing the cloud at the edge of the RE beam. Within the RE density ramp, the pellet cloud evolution strongly depends on the impact ionization by REs which is crucial for causing strong MHD effects. Both 3D and 1D simulations showed that the longitudinal heat diffusion provides negligibly small factor to  the total energy balance. Single fragment simulations quantified various terms in the transverse dynamics model. The penetration depth of the ablated cloud into the RE region was about 55 cm, reached at 1.0 ms. At about 3 ms , the cloud reconnected longitudinally at q=2.

SPI simulations show details of the evolution of ablated clouds, in particular the expansion along magnetic field lines, cross-field transport due to grad-B drift, Alfven wave drag and connection current damping terms, global dynamics of the ablated cloud at tokamak scales as well as the maximum penetration of the leading edge of ablated clouds into the RE region. We observe that a fraction of the ablated hydrogen gas (10 - 15\%) is expelled from the RE beam by pressure gradients soon after each fragment injection. The expelled cold gas slowly expands in the background plasma. In the RE beam, the ablated material becomes ionized by both the thermal Saha process and by direct RE impact, absorbs energy of RE's, rapidly propagates along magnetic field lines and experiences various cross-field drift effects. The SPI cloud penetrates approximately 40 cm into the RE beam for the given RE ramp profile. The deepest penetration was achieved at 1.1 - 1.2ms. For $q=2$, the cloud reconnects longitudinally at 1.7ms. At these stages of the cloud evolution, more than 50\% of the ablated gas is moved outside of the RE beam by the grad-B drift. 

We also studied the dependence of ablation cloud evolution processes on the number of fragments in the SPI plume. More material was initially deposited into the RE beam by the SPI of 5000 fragments compared to 555 fragments: for the case of smaller number of large fragments, more neutral gas was expelled from the RE region at the initial stage of ablation. But depositing more material into the RE beam led to higher ablation cloud pressure values and stronger grad-B drift. As a result, a larger fraction of hot ablated material drifted outside the RE beam in the case of 5000 fragment SPI. Despite these differences, the global dynamics and the cloud reconnection time were similar. In particular, both clouds achieved very similar penetration depth at any time.  

In the future, we will improve physics models for REs. In particular, we will add an evolution model for the RE beam that accounts for REs decay in  the ablated material.

\vskip5mm
{\bf Acknowledgments}
This research has been partially supported by the U.S. Department of Energy grants SciDAC Center for Tokamak Transient Simulations and DE-FG02-07ER54917. The authors are grateful to N. Garland for providing electron impact ionization rates. Permission to use PrismSPECT from I. Golovkin is gratefully acknowledged.

\vskip5mm
{\bf Data availability}
The data that support the findings of this study are available
from the corresponding author upon reasonable request.

\vskip5mm

\bibliographystyle{plain}
\bibliography{ref} 

\begin{thebibliography}{10}

\bibitem{BaileyRochau2009}
J.~E. Bailey, G.~A. Rochau, R.~C. Mancini, C.~A. Iglesias, J.~J. MacFarlane,
  I.~E. Golovkin, C.~Blancard, Ph. Cosse, and G.~Faussurier.
\newblock Experimental investigation of opacity models for stellar interior,
  inertial fusion, and high energy density plasmas.
\newblock {\em Physics of Plasmas}, 16(5):058101, 2009.

\bibitem{Baylor_2000}
L.R. Baylor, T.C. Jernigan, P.~Gohil, G.L. Schmidt, K.~H. Burrell, S.K. Combs,
  D.R. Ernst, C.M. Greenfield, R.J. Groebnerand, W.A. Houlberg, et~al.
\newblock Improved fueling and transport barrier formation with pellet
  injection from different locations on {DIII-D}.
\newblock {\em in Proc. 18th Fusion Energy Conference, IAEA-CN-77}, 01 2000.

\bibitem{BosvielParks2021}
Nicolas Bosviel, Paul Parks, and Roman Samulyak.
\newblock Near-field models and simulations of pellet ablation in tokamaks.
\newblock {\em Physics of Plasmas}, 28(1):012506, 2021.

\bibitem{breizman2019physics}
B.N. Breizman, P.~Aleynikov, E.~M. Hollmann, and M.~Lehnen.
\newblock Physics of runaway electrons in tokamaks.
\newblock {\em Nuclear Fusion}, 59(8):083001, 2019.

\bibitem{BursteddeWilcoxGhattas11}
Carsten Burstedde, Lucas~C. Wilcox, and Omar Ghattas.
\newblock {\texttt{p4est}}: Scalable algorithms for parallel adaptive mesh
  refinement on forests of octrees.
\newblock {\em SIAM Journal on Scientific Computing}, 33(3):1103--1133, 2011.

\bibitem{Commaux2010}
N.~Commaux, B.~P\'egouri\'e, L.R. Baylor, F.~Kochl, P.B. Parks, T.C. Jernigan,
  A.~G\'eraud, and H.~Nehme.
\newblock Influence of the low order rational $q$ surfaces on the pellet
  deposition profile.
\newblock {\em Nucl. Fusion}, 50:025011, 2010.

\bibitem{NGarland}
Nathan Garland.
\newblock Private communication.

\bibitem{HollmannBortolon2022}
E.~M. Hollmann, A.~Bortolon, F.~Effenberg, N.~Eidietis, D.~Shiraki, I.~Bykov,
  B.~E. Chapman, J.~Chen, S.~Haskey, J.~Herfindal, A.~Lvovskiy, C.~Marini,
  A.~McLean, T.~O'Gorman, M.~D. Pandya, et~al.
\newblock Dynamic measurement of impurity ion transport in runaway electron
  plateaus in {DIII-D}.
\newblock {\em Physics of Plasmas}, 29(2):022503, 2022.

\bibitem{HollmannBykov2020}
E.M. Hollmann, I.~Bykov, N.W. Eidietis, J.~L. Herfindal, A.~Lvovskiy, R.A.
  Moyer, P.B. Parks, C.~Paz-Soldan, , A.~Yu. Pigarov, D.L. Rudakov, et~al.
\newblock Study of argon expulsion from the post-disruption runaway electron
  plateau following low-{Z} massive gas injection in {DIII-D}.
\newblock {\em Phys. Plasmas}, 27:042515, 2020.

\bibitem{HollmannNaitlho2022}
E.M. Hollmann, N.~Naitlho, S.~Yuan, R.~Samulyak, P.B. Parks, D.~Shiraki,
  J.~Herfindal, and C.~Marini.
\newblock Measurement and simulation of small cryogenic neon pellet {Ne-I} 640
  nm photon efficiency during ablation in {DIII-D} plasma.
\newblock {\em Phys. Plasmas}, 2022.
\newblock Submitted.

\bibitem{Kiramov2020}
D.I. Kiramov and B.~N. Breizman.
\newblock Pellet sublimation and expansion under runaway electron flux.
\newblock {\em Nuclear Fusion}, 60(8):084004, 2020.

\bibitem{Koechl2012}
F.~Koechl, B.~P\'egouri\'e, A~Matsuyama, H.~Nehme, D.~Frigione, L.~Garzotti,
  G.~Kamelander, and V.~Parail.
\newblock Modelling of pellet particle ablation and deposition: The hydrogen
  pellet injection code {HPI2}.
\newblock EFDA–JET–PR(12)57:1--86, 2012.

\bibitem{Matsuyama2012}
A.~Matsuyama, F.~Koechl, B.~P\'egouri\'e, R.~Sakamoto, G.~Motojima, and
  H.~Yamada.
\newblock Modeling of the pellet depostion profile and $\nabla${B}-induced
  drift displacament in non-axisymmetric configurations.
\newblock {\em Nucl. Fusion}, 52:123017, 2012.

\bibitem{Parks_2005}
P.B. Parks and L.R. Baylor.
\newblock Effect of parallel flows and toroidicity on cross-field transport of
  pellet ablation matter in tokamak plasmas.
\newblock {\em Physical Review Letters}, 94:125002, 2005.

\bibitem{Paz-Soldan2021}
C.~Paz-Soldan, C.~Reux, K.~Aleynikova, P.~Aleynikov, V.~Bandaru, M.~Beidler,
  N.~Eidietis, Y.Q. Liu, C.~Liu, A.~Lvovskiy, et~al.
\newblock A novel path to runaway electron mitigation via deuterium injection
  and current-driven {MHD} instability.
\newblock {\em Nuclear Fusion}, 61(11):116058, oct 2021.

\bibitem{ReuxPaz-Soldan2021}
C\'edric Reux, Carlos Paz-Soldan, Pavel Aleynikov, Vinodh Bandaru, Ondrej
  Ficker, Scott Silburn, Matthias Hoelzl, Stefan Jachmich, Nicholas Eidietis,
  Michael Lehnen, et~al.
\newblock Demonstration of safe termination of megaampere relativistic electron
  beams in tokamaks.
\newblock {\em Phys. Rev. Lett.}, 126:175001, Apr 2021.

\bibitem{SamulyakLu2007}
R.~Samulyak, T.~Lu, and P.B. Parks.
\newblock A magnetohydrodynamic simulation of pellet ablation in the
  electrostatic approximation.
\newblock {\em Nuclear Fusion}, 47(2):103--118, Jan 2007.

\bibitem{SamWangChen2018}
R.~Samulyak, W.~Xingyu, and H.-C. Chen.
\newblock {Lagrangian particle method for compressible fluid dynamics}.
\newblock {\em Journal of Computational Physics}, 362:1--19, June 2018.

\bibitem{SamulyakYuan2021}
R.~Samulyak, S.~Yuan, N.~Naitlho, and P.~B. Parks.
\newblock Lagrangian particle model for 3d simulation of pellets and spi
  fragments in tokamaks.
\newblock {\em Nuclear Fusion}, 61(4):046007, 2021.

\bibitem{Shiraki2018}
D.~Shiraki, N.~Commaux, L.R. Baylor, C.M. Cooper, N.W. Eidietis, E.M. Hollmann,
  C.~Paz-Soldan, S.K. Combs, and S.J. Meitner.
\newblock Dissipation of post-disruption runaway electron plateaus by shattered
  pellet injection in {DIII}-d.
\newblock {\em Nuclear Fusion}, 58(5):056006, mar 2018.

\bibitem{Zeldovich}
Y.B. Zel'dovich and Y.P. Raiser.
\newblock {\em Physics of shock waves and high temperature hydrodynamic
  phenomena}.
\newblock Dover, 2002.

\end{thebibliography}

\end{document}